\documentclass[aps,prd,showpacs,nofootinbib,amsmath,amssymb,mcite,12pt,preprint,a4paper]{revtex4-1}
\pdfoutput=1

\usepackage{amsfonts,amsmath,amssymb,amsthm}

\usepackage{graphicx}
\usepackage{slashed}
\usepackage[colorlinks]{hyperref}

\usepackage{varioref,exscale,latexsym,amsmath,amssymb}
\usepackage{graphicx}

\usepackage{graphicx}
\usepackage{slashed}
\usepackage{dcolumn}
\usepackage{bm}

\def\beq{\begin{equation}}

\def\eeq{\end{equation}}

\def\beqa{\begin{eqnarray}}

\def\eeqa{\end{eqnarray}}

\begin{document}
\preprint{ACFI-T19-12}

\title{{\bf A Critique of the Asymptotic Safety Program}}

\medskip\

\author{ John F. Donoghue${}$}
\email[Email: ]{donoghue@physics.umass.edu}
\affiliation{~\\
Department of Physics,
University of Massachusetts\\
Amherst, MA  01003, USA\\
 }

\begin{abstract}
The present practice of Asymptotic Safety in gravity is in conflict with explicit calculations in low energy quantum gravity. This raises the question of whether the present practice meets the Weinberg condition for Asymptotic Safety. I argue, with examples, that the running of $\Lambda$ and $G$ found in Asymptotic Safety are not realized in the real world, with reasons which are relatively simple to understand. A comparison/contrast with quadratic gravity is also given, which suggests a few obstacles that must be overcome before the Lorentzian version of the theory is well behaved. I make a suggestion on how a Lorentzian version of Asymptotic Safety could potentially solve these problems.
\end{abstract}
\maketitle

\tableofcontents

\section{Preface}

Asymptotic freedom describes the situation where the coupling constants of a quantum field theory run to zero at asymptotically high energy. For renormalizeable theories, this running is logarithmic in the momentum.

Asymptotic safety (AS) describes the situation where the coupling constants run to an ultraviolet fixed point where the couplings are finite but where the beta functions vanish. While this can happen in a renormalizeable field theory \cite{Litim:2014uca} where the running is logarithmic, its most common application is in the study of gravity \cite{Weinberg, Niedermaier:2006wt, Reuter:1996cp, Reuter:2019byg} . In this case, the running is generically power-law, because of the dimensional coupling constants. In this paper I am discussing only the gravitational case with power-law running.

There is a conflict between the much of the present practice in AS and known explicit calculations of quantum processes in quantum gravity. This was originally pointed out in work with M. Anber \cite{Anber:2011ut}. At low energy calculations of quantum gravity processes can be carried out in the rigorous Effective Field Theory (EFT) treatment \cite{Donoghue:1994dn, Donoghue:2017pgk} and we can compare these observables with the practice of Asymptotic Safety.  {The EFT is valid at low energies, which in this case means below the Planck scale. The major action in Asymptotic Safety happens around the Planck scale. Nevertheless, the AS techniques also apply below this scale, and predictions only emerge by running the cutoff to zero energy. Therefore in the overlap region we can make this comparison.} More recently, explorations of quadratic gravity \cite{Stelle:1976gc, Julve:1978xn,  Fradkin:1981iu, Avramidi:1985ki, Salvio:2018crh, Salvio:2014soa, Donoghue:2018izj, Donoghue:2019fcb, Donoghue:2019ecz, Einhorn:2014gfa, Holdom},  which involves curvature-squared terms in the action, also shed light on the connection to AS. Quadratic gravity is a renormalizeable theory for quantum gravity in the ultraviolet. It is somewhat more tentative and needs further exploration itself. However, it provides a calculational framework which is reasonably close to AS, such that it provides an interesting lessons for AS.

The present paper is an attempt to explain many of the issues involved. It has been invited to be part of a volume describing an overview of running couplings in gravity. It is meant both as a summary of concerns aimed at the AS community, and as an explication of the core issues for an outsider audience. As such it will contain comments which are unnecessary for an AS practitioner, as well as occasional technical details aimed only at the experts. I hope that this document can serve this dual purpose.

The reader will also notice that I often use the phrase ``present AS practice''. This is because I want to differentiate between what is often done in the present AS literature from what could be the ultimate understanding of Asymptotic Safety. The AS paradigm is potentially an attractive resolution to the puzzle of quantum gravity. However, the present status is not yet a successful resolution. This article is then an attempt to point out shortcomings in the present practice as well as to point to future directions which may be fruitful.

\subsection{Key contrasts: Euclidean vs Lorentzian, powers vs logarithms, cutoffs vs dimensional regularization}

As a preview to the more technical discussion which follows, let me mention some of the important issues which are central to that discussion.

The foundational technique of AS practice is the Euclidean functional integral. One studies this with an infrared cutoff and integrates out quantum effect in an energy scale around the cutoff. This is a variation of our usual way of using cutoffs in that the cutoff is introduced to keep the quantum effects {\em above} the cutoff and removes those with scales below the cutoff. The variation of the coupling parameters with that scale gives the renormalization group flow of the couplings. It is understood that running the cutoff down from the UV fixed point down to a zero value for the cutoff will then include all of the quantum corrections.

However, it is also common practice in the community to assign a meaning to the parameters at given values of the cutoff. For example, the running Newton constant in AS is often parameterized as
\beq\label{runningG}
G(k) = \frac{G}{1+ Gk^2/g_*}
\eeq
where $k$ is the cutoff, and $g_*$ is related to the fixed point in a way that will be described below. The use of the symbol $k$ makes it tempting to think of $k$ as a momentum (in practice it is closer to a mass cutoff) and to think of the resulting $G(k)$ as one that depends on the momentum scales in a reaction. This is incorrect, as we will see from direct examples in Sect. \ref{explicit}. Moreover, even if it were a Euclidean momentum, its Lorentzian counterpart would be ill-defined. A large Euclidean momentum can translate to a massless on-shell Lorentzian particle if $k_0^2-\mathbf{k}^2=0$ or to positive or negative values of the various kinematic invariants in reactions (i.e. $s>0$ or $t<0$) The basic question then is whether $G(k)$ at finite values of the cutoff has any physical meaning. Explicit calculations suggest that it does not.

A second point to watch is that the important features of AS do not occur when dimensional regularization is used. For example, if one truncates to the Einstein action, then the Newton constant does not run in dimensional regularization, contradicting Eq. \ref{runningG}. At one level, this can be blamed on a known weakness of dimensional regularization. Near $d=4$ it cannot identify quadratic divergences as it includes integrations over all scales. So it is perfectly allowable to use cutoffs to identify effects at a particular scale around the cutoff. But in the end, {\em real physics should not depend on the regularization scheme}. I take it as given that dimensional regularization provides an acceptable regularization scheme to describe physical processes in field theory. I know of no counter-example. {Moreover, I use dimensional regularization in the perturbative regime where it use in scattering amplitudes is unquestioned.} So in the end, any scheme which uses cutoffs to define the theory should give the same physical predictions {in such reactions}. We need to understand how AS can do that. This is not a trivial constraint. In fact, we can understand how this occurs, but the resolution tells us that the running $G(k)$ is not valid for physical processes.

The other feature to be aware of, before we start describing the details, is the difference between logarithmic running constants and power-law running. Our experience in renormalizeable field theories is with logarithmic running. The need to use running couplings comes from the existence of large logarithms. If we measure the coupling at a renormalization scale $\mu_r$ and apply it at an energy scale $s$, there will be large corrections of order $\alpha({\mu_r}) \log (s/\mu^2_r)$. Use of the renormalization group lets us take that original measurement up to the scale $\mu_r^2\sim ~s, ~t$, in which case there are no longer any large logarithms. Note that the signature of the kinematic invariants does not matter as $\log s/\mu_r^2 \sim \log t/\mu_r^2$ up to small factors as long as $s$ and $t$ are both of order $\mu_r^2$, even though $s$ and $t$ have opposite signs. Moreover, $\mu_r $ is an unphysical parameter. In the end, $\mu_r$ disappears from physical processes.

However, AS applied to gravity requires something different, which is power-law running. Because most of the couplings in the most general Lagrangian are dimensionful, one multiplies them by powers of the scale in order to define dimensionless variables. For example the Newton constant is modified by
\beq
g(k)= Gk^2
\eeq
The running of this dimensionless coupling is that which defines the fixed point. In this case
\beq
g \underset{k \to \infty}{\to} = g_* \ \ ,
\eeq
hence the notation of Eq. \ref{runningG}. However, now we must make contact with physical processes. If we imagine measuring $G$ at some scale $\mu_r^2$, one is faced with the question of making the measurement of at some values of $s$ or $t$ of order $\mu_r^2$. But $s$ and $t$ generally carry opposite signs, and $g(s)$ and $g(t)$ are wildly different quantities in a way that does not occur in logarithmic running. Moreover, as we will see, there is no reason to expect that something like $G(s)$ captures the actual effect of quantum corrections to $G$. Higher order momentum dependence generally refers to new operators, where the factors of $s$ or $t$ come from extra derivatives on the fields. These new operators need not enter reactions in the same way as the lowest order operator.

\section{Foundational issues}

{In this section, we discuss several issues associated with running coupling constants. Therefore, let me be clear what I mean by a running coupling constant. It is a coupling defined to depend on a scale which captures essential quantum corrections in physical processes for physics around that scale. The fact that it is useful in physical processes is important. We will see that this aspect is also part of the original formulation of Asymptotic Safety by Weinberg \cite{Weinberg}. A useful running coupling should also apply to more than one process - it should be universally valid. If there is a scale dependence in some function which however does not have a direct physical meaning, we do not refer to this as a running coupling.  }

\subsection{There is no gravitational running of regular coupling constants}\label{couplingconstants}

There are obviously gravitational corrections to ordinary reactions which occur in the Standard Model. Robinson and Wilczek suggested that it could be useful to define the gravitational correction to the running coupling constants of the theory \cite{Robinson:2005fj}. For example, for the gauge couplings, this could take the form
\beq
\beta(g, E) \equiv \frac{d g}{d \ln E}=-\frac{b_{0}}{(4 \pi)^{2}} g^{3}+a_{0} \frac{E^{2}}{M_{\mathrm{P}}^{2}} g
\eeq
After a large number of papers in the literature \cite{doesnot, Anber:2010uj, does2, Toms:2011zz}, on various sides of this issue, it has become clear that this does not occur. The reasons are instructive for our discussion of Asymptotic Safety.

The first significant reason is kinematic. In Lorentzian reactions, the variable $E^2$, can have either a positive or negative sign. For example, if the reaction $e^+e^- \to \mu^+ \mu^-$ has the gravitational correction
\beq
{\cal M}\sim \frac{e^2 (1-aGs)}{s}
\eeq
where $s=(p_1+p_2)^2>0$ and $a$ is some constant. For the reaction $e^+\mu^- \to e^+\mu^-$, related to it by crossing symmetry, will have the form
\beq
{\cal M} \sim \frac{e^2 (1-aGt)}{t}
\eeq
with $t=(p_1 -p_3)^2 <0$ having the opposite sign from $s$.
The gravitational corrections will go in different directions in the two reactions. If the first reaction has a decreasing coupling, the second one will have an increasing coupling. In more complicated QED reactions, there will be many kinematic invariants which span the range of sizes and signs. These effects cannot be captured by a running coupling constant. If one attempts to measure the effective electric charge at a renormalization scale $s=\mu_r^2$ using $e^+e^- \to \mu^+ \mu^-$, such as $e^2(\mu_R)= e^2 (1-aG\mu_r^2)$ that coupling will not be useful in describing the crossed reaction or in other more complicated reactions.

The other significant reason is universality. The gravitational corrections carrying powers of the energy are not actually a renormalization of the electric charge, but are described by new operators with extra derivatives. For example, if we take the bare QED Lagrangian to be
\beq
{\cal L} = \frac{1}{4e_0^2} F_{\mu\nu}F^{\mu\nu}
\eeq
then after loop corrections the energy dependent terms would be reflected in operators such as
\beq
{\cal L} = \frac{1}{4e^2} F_{\mu\nu}F^{\mu\nu}+ a G F_{\mu\nu}\Box F^{\mu\nu} + b G\bar{\psi} \sigma_{\mu\nu} i\slashed{D} \psi \partial_\mu A_\nu + c G \bar{\psi}  i\slashed{D} D^2 \psi +...
\eeq
These operators can enter different reactions in different ways, depending on the particle content and kinematics of those processes. Their contribution is not generally in the same manner as the original renormalized charge, and then is not generally able to be described by a running charge.

It should be noted that because the graviton is massless, not all the gravitational corrections are described by local operators. There can be non-local effects reflecting the long distance propagation of the graviton. However, this feature does not change the discussion above.

This brief discussion follows most closely Ref. \cite{Anber:2010uj} where further examples are given, but is also reflected in different ways in Refs. \cite{doesnot}.

\subsubsection{Using a cutoff does not imply the running of a coupling constant}

In response to criticisms such as the above, some authors suggested that using a cutoff regularization scheme would produce a running coupling \cite{does2}. This is not correct, and again it is useful for our purposes to understand why.

We first note that using dimensional regularization there is no gravitational renormalization of the electric charge when neglecting the masses of the fermions. This follows from power-counting with a dimensional coupling $G$. Temporarily neglecting the fermion masses, the only dimensional factor in dimensional regularization comes from the factor $\mu^{4-d}$ inserted in Feynman integrals in order to keep the dimensions correct. This yields factors of $\log \mu^2$ in intermediate steps in calculations but could never produce a factor $G\mu^2$ in gravitational calculations. With fermion masses, the gravitational corrections are of the form
\beq
{\cal L} = \frac{1+a Gm^2}{4e_0^2} F_{\mu\nu}F^{\mu\nu} +...
\eeq
where $a$ is some constant and the ellipses refer to the momentum dependent corrections discussed above. When measuring the electric charge one finds
\beq
\frac{1+a Gm^2}{4e_0^2} =\frac{1}{4e_r^2}
\eeq
and one expresses predictions in terms of the renormalized charge $e_r$. One is left only with the momentum dependent operators described above.

Real physics does not depend on the nature of the regularization scheme. However, the authors \cite{does2} suggested that the use of a cutoff regularization could be used to define a running coupling which would capture the quantum gravitational effects at a given scale. That is, by using a cutoff $\Lambda$ one would define the beta function
\beq
\beta(g, \Lambda) \equiv \frac{d g}{d \ln \Lambda}=-\frac{b_{0}}{(4 \pi)^{2}} g^{3}+a_{0} \frac{\Lambda^{2}}{M_{\mathrm{P}}^{2}} g
\eeq
This would get around the kinematic and universality problems of the Robinson-Wilczek suggestion.
The reasoning is vaguely Wilsonian - by using a cutoff one includes effects which occur below that scale. One rebuttal is that one must also include effects which occur above that scale, and the overall physics is independent of the separation scale. However, even if one neglects this, the cutoff effect disappears in renormalization procedure. The introduction of a cutoff does lead to a renormalization of the bare electric charge, of the form
\beq
{\cal L} = \frac{1+a_0 G\Lambda^2}{4e_0^2} F_{\mu\nu}F^{\mu\nu} +....
\eeq
with the suggestion that
\beq
\frac{1+a_0 G\Lambda^2}{4e_0^2} = \frac{1}{4e(\Lambda)^2}
\eeq
However when one calculates a physical process, this effect enters the amplitude just like the renormalized charge, and the correct identification is
\beq
\frac{1+a G\Lambda^2}{4e_0^2} =\frac{1}{4e_r^2}
\eeq
and this manifestation of $\Lambda$ disappears from the physical amplitude \cite{Toms:2011zz}. In the end, cutoff regularization and dimensional regularization do agree in physical amplitudes.

{Here we have seen the definition of a coupling constant which depends on a scale - the cutoff $\Lambda$. In that sense it is a truism that it ``runs''. However, it does not qualify as a ``running coupling constant'' because that running is not relevant for physical processes at energies around that scale. Indeed the cutoff dependence is completely unphysical - it disappears from all amplitudes. If we wish to describe its scale dependence we should come up with a different name for it. Perhaps ``incomplete coupling constant'' is appropriate, as it is defined to include only quantum corrections below the cutoff scale. When used as a UV regulator, we do not care about the incompleteness, as the true physics beyond is unknown and in any case irrelevant for low energy processes. But if we are trying to use the cutoff as a running parameter at the scale of the energies being studied, we do care about the incompleteness. The full coupling constant does not have such a division.}

\subsubsection{Log running vs power-law running}

The above sections illustrate a truism - {\em There are no power-law running coupling constants in 4D Minkowski quantum field theory}.

Logarithmic running works because the logarithm is directly tied to renormalizaton. In the QED case, photon exchange with the vacuum polarization leads to a factor of
\beq
{\cal M} \sim \frac{e_0^2}{q^2[1+ \Sigma(q)]+i\epsilon}
\eeq
where $\Sigma(q)$ is scalar part of the vacuum polarization. No matter how one chooses to regularize it, the vacuum polarization contains a divergent term and a logarithm of $q^2$. The divergence and the logarithm share the same coefficient. If we measure the charge using $e^+e^- \to \mu^+ \mu^-$  at a renormalization scale $s=\mu_R^2$ with $s=(p_1+p_2)^2>>m_e^2$, this result becomes
\beq
\frac{e^2(\mu_R)}{s[1-\frac{\alpha}{3\pi}\log \frac{-s}{\mu_R^2}]+i\epsilon}
\eeq
Because the logarithm comes along with charge renormalization, it occurs in every reaction in the same fashion. And because of the properties of the logarithm, the same running coupling would apply to the crossed reaction  $e^+\mu^- \to e^+ \mu^-$ with the change $s\to t$.

Power-law effects do not share these features. There is no universal connection of power-law corrections to the renormalization of the charge. And because of Minkowski kinematics, the effects in different channels can go in opposite directions.

That being said, it is possible in any one calculation to define a running coupling {\em for that particular process}. This may be a useful procedure. However, in field theory, a coupling constant has multiple duties. It not only describes that one particular process, but also must describe a multitude of others. These can differ in the arguments, i.e. $\lambda (\phi)$ vs $\lambda(q^2)$, and also on the nature of the process. The same coupling needs to describe not only space-like vs time-like reactions such as we have used as examples above, but also multi-particle reactions which involve many more particles than the simplest reaction. It is this multiplicity of uses where attempts to define power-law running couplings fail. The same definition which works in one setting will in general fail in the these other settings. The logic and mathematics which tell us that logarithmic running coupling constants are useful does not apply to power-law running.

The reader may object that Wilson has taught us the value of coarse-graining as a way to define couplings at different scales, and that this procedure has been verified in condensed matter systems even including power-law re-scalings. However, the couplings in these condensed matter examples do not have as many applications as the couplings in scattering processes. And the 3D setting for condensed matter systems does not display the kinematic variety of Minkowski reactions. It is easy to understand how the Wilsonian rescaling in condensed matter may be useful, while corresponding Minkowski QFT applications are more complicated.

\subsection{Weinberg formulation of Asymptotic Safety}

The vision for Asymptotic Safety for gravity was formulated by Weinberg \cite{Weinberg}. He invokes a situation where all the coupling constants run to fixed values at high energy. This includes the dimensionful couplings, when rescaled by a universal dimension. He defines dimensionless variables $g_i$ by multiplying by a scale $\mu$. For example, one would have $g_G = G\mu^2$ and $g_\Lambda = \Lambda_{vac}/\mu^4$, where $\Lambda_{vac}$ is the vacuum energy density\footnote{I will try to keep separate the vacuum energy density $\Lambda_{vac}$ (which much of the particle physics community refers to as the cosmological constant) from other definitions of the cosmological constant. Much of the Asymptotic Safety community uses the symbol $\Lambda$ for a different version $\Lambda=  -\Lambda_{vac}/8\pi G = -\Lambda_{red}$. For this combination, I will use $\Lambda_{red}$ (with $red$ standing for ``reduced'')}.

Specifically, {in his 1979 paper \cite{Weinberg}} Weinberg formulates the hypothesis using scattering processes and other reactions. Using these dimensionless coupling he suggests that these rates could have the form
\beq\label{initial}
R = \mu_R^D f\left[\frac{E}{\mu_R}, X, g_i(\mu_R)\right]
\eeq
where $X$ stands for all the other dimensionless physical variables. Here $\mu_R$ is meant to be a renormalization point, as used above. Because physics cannot depend on the arbitrary choice of the renormalization point, one can choose $\mu_R = E$ and have the result that the rate behaves as
\beq\label{rates}
R = E^D f\left[1, X, g_i(E)\right]
\eeq
Aside from the pre-factor (which would involve $D=-2$ for a total cross-section) the rates would then depend on the couplings $g_i(E)$ as $E\to \infty$. Asymptotic safety is defined by the condition that the running couplings go to constant values $g_i(E)\to g_{i*}$ at high energy, or equivalently that their beta functions vanish
\beq\label{fixed}
\beta(g_i) =E \frac{\partial}{\partial E}g_i = 0
\eeq
This is the UV fixed point. The implication here is that instead of blowing up with the energy, as $GE^2$ would, these factors go to constant values. {I will refer to Eqs. \ref{initial}-\ref{fixed} as the Weinberg conditions for Asymptotic Safety. }

We can see from the discussion of coupling constants in the previous subsection that this needs to be generalized somewhat, as there is no unique energy $E$ in Minkowski reactions. We do not want to include the kinematic variables in the running parameters, such as $g_i(s),~g_i(t),...$ because of the kinematic ambiguity and differing signs. The best that we can hope for is to choose all of the kinematic variables of order the renormalization point, $|s|\sim |t|\sim ...\sim \mu_R^2$ and write the rate as
\beq
R = \mu_R^D f\left[\frac{s}{\mu_R^2},\frac{t}{\mu_R^2},... X, g_i(\mu_R)\right]  \ \ .
\eeq
In this formulation it is not clear how the renormalization scale $\mu_R$ drops out of physical observables. However, that can work out {\em in a given process} by explicitly performing the renormalization and demanding that the result is independent of $\mu_R$. That demand then identifies the renormalization group flow of the couplings. The larger question is whether, having done this renormalization in one process, the result generalizes to other processes and is useful in describing the quantum effects of the full theory. {This raises the possibility that the Weinberg conditions themselves are unworkable when applied to a full set of reactions with many kinematic variables of differing signs. Our comparison with explicit reactions below will be discouraging in this respect when applied to $G$ and $\Lambda$. However, if Asymptotic Safety is to be successful there must be a modified version of these conditions which applies for the high energy limit of physical processes. I will continue to use the Weinberg formulation as the vision for the AS program.}

{In our discussion of the present practice of Asymptotic Safety, it is important to point out that the Weinberg proposal is for true running couplings in the sense that we are using that phrase in this paper. That is, these are couplings that apply in physical reactions (in particular as functions of energy) and which in a useful way capture relevant quantum corrections appropriate for those energies. }

\subsection{The practice of Asymptotic Safety}

This section is clearly meant primarily for readers outside the AS community. It tries to very briefly explain the formalism and physics of the calculations. However, there are important comments towards the end of Sect. \ref{oneloop} that are intended for all readers.

The present practice of Asymptotic Safety does not study reaction rates, but rather evaluates the flow of the Euclidean functional integral in a background field formulation { - the Euclidean functional renormalization group (FRG)}. That is, the functional integral is a function of the metric, curvatures and covariant derivatives.. The logic here is that once all quantum corrections are included in the Euclidean functional integral, the result can be continued to Lorentzian spaces, and the metric and curvatures expanded in the external fields in order to obtain the amplitudes that the Weinberg criterion envisions. I will call this the {``ideal FRG program''}.

However, for the most part in present applications this logic is not followed in practice\footnote{Codello et al. \cite{Codello:2015oqa} have pursued the ideal {FRG} program to reproduce some of the results of chiral perturbation theory. The chiral logs emerge in the IR limit as $k\to 0$}. Rather rather than evaluating the full functional integral, one evaluate the evolution from the UV fixed point down to some cutoff $k$ including quantum corrections above $k$. Without evaluating the quantum corrections below the cutoff, it is then assumed that the resulting $g_i(k)$ are the appropriate couplings to use in something like the Weinberg criterion in real world applications at the scale $k$. That is, $g_i(k)\sim g_i(E\sim\mu_R\sim k)$. There is also necessarily a truncation of the basis (to be discussed soon) in such applications. There is an extra logical step required if these assumptions are to be true.  This can be called the practical AS program.

One complication of the AS program is that the basis set of operators is infinite, with a corresponding infinite number of coupling constants. The renormalization flow for a theory such as gravity mixes operators of all dimensions, with the only restriction being that of general covariance. In the action, there will be local terms of the form
\beq\label{generalL}
{\cal L}= \sqrt{-g}\left[-\Lambda_{vac} - \frac{1}{16\pi G}R + c_1 R^2 + c_2 C_{\mu\nu\alpha\eta}C^{\mu\nu\alpha\eta}+ d_1 R^3+d_2 R\Box R+...\right]
\eeq
This series can be ordered by powers of derivatives, such that only the operators with few derivatives are relevant for the low energy limit. This is what is done in the effective field theory treatment. However, Asymptotic Safety concerns the high energy limit and all operators become active as the energy goes to infinity. The ideal {FRG} program then would involve all possible operators with their coefficients\footnote{There are also non-local contributions to the functional integral. It is assumed that these are fully parameterized by the coefficients of the local operators.}.  However in the ideal {FRG} program these coefficients are not all independent. The infinite set of couplings would be described by a few {\em relevant} couplings and only special values of the parameters would be consistent with the Asymptotic Safety hypothesis.

Practicality requires that this be truncated at some order. The AS community has explored a remarkable range of such truncations, and the overall picture that emerges has so far been independent of the truncation. For the purposes of this paper, I will assume that the truncation problem is not a fundamental obstacle. Nevertheless, we can examine truncations to see what might be issues for the full program, as in Sects. \ref{tachyons} and \ref{obstacles}.

The fundamental equation of AS practice, the Wetterich equation \cite{Wetterich:1992yh}, describes the change of the Euclidean functional integral $\Gamma_k$, again defined to include quantum fluctuations above the scale $k$, under a change in scale\footnote{The Wetterich equation is more general than its application to AS, and Asymptotic Safety could in principle be addressed without the Wetterich equation (i.e. see Sec. \ref{alternative} for a possiblity). However, present practice in AS involves this equation.} .
\beq
k \frac{\partial}{\partial k}\Gamma_k = \frac12 {\rm Tr}\left[\left(\frac{1}{\frac{\delta^2 \Gamma_k }{\delta g \delta g}+R_k}\right)~k\frac{\partial}{\partial k} R_k~\right]
\eeq
Here $R_k$ is the cutoff function which suppresses momentum modes below $k$. Conceptually, it is like a mass below the scale $k$ and zero above $k$, chosen in some smooth way so that there is not a discontinuity.  An example is
\beq
R_k = (k^2-D^2)\theta (k^2-D^2)
\eeq
In understanding the variation $\delta^2 \Gamma_k /\delta g \delta g$, one notes that $g$ schematically represents the metric and any other fields in the theory. If the functional contained $D_\mu g D^\mu g$ then the variation would be $-D^2$. So conceptually, this equation is similar to $k\partial_k {\rm Tr} \log (D^2 +m^2_k)$. Of course the real case is very much more complicated by the interactions and all the indices. A positive feature of the flow equation is that the flow only depends on the physics near the cutoff scale $k$. Higher scales have already been included and no longer enter because of the vanishing of $\partial_k R_k$ at high $k$, while lower scales are suppressed by the cutoff. Qualitative results have so far been independent of the choice of the function, although numerical results do depend modestly on the choice.

Weinberg in his Erice lectures on critical phenomena \cite{Weinberg:1976xy} also expressed a similar structure for the running coupling.

Much work has gone into exploring the existence and properties of the UV fixed points. To do this one first identifies a truncation in the basis. One starts at finite $k$ and uses the Wetterich equation to flow to higher scales. In the infinite dimensional space of coupling constants, the fixed points live on finite dimensional ``critical surface''. Common expectation is that this is two or three dimensional. This leaves a two or three dimensional family of solutions. When one flows from the fixed point to the IR at $k=0$, one will have two or three undetermined constants. In particular $\Lambda_{vac}$ and $G$ at $k=0$ are not predicted. But in principle there are predictions for an infinite number of other constants in the local effective Lagrangian.

\subsubsection{AS at one-loop}\label{oneloop}

In order to see the FRG machinery at work, we can look at the illuminating calculation of Codello and Percacci \cite{Codello:2006in}, which is described as a one-loop evaluation including terms up to the order curvature-squared. This example also allows a comparison with a conventional treatment of quadratic gravity, which will be given in Sec. \ref{quadratic}.

The Euclidean action is parameterized by five couplings, in the form

\beq
S= \int d^{4} x \sqrt{g}\left[\frac{1}{8\pi G} \Lambda_{red}-\frac{1}{16\pi G} R+\frac{1}{2 \lambda} C^{2}-\frac{\omega}{3 \lambda} R^{2}+\frac{\theta}{\lambda} E\right] \ \  .
\eeq
Here $C^2$ is the Weyl tensor squared, and $E$ is the Gauss-Bonnet term. The vacuum energy is defined by $\Lambda_{vac}= -\frac{1}{8\pi G} \Lambda_{red}$. In four dimensions, $E$ is a total derivative and does not influence any local physics. This will be evidenced in the flow as the parameter $\theta$ does not influence any of the other physical parameters. The dimensionful parameters are $\Lambda_{red}$ and $G$, while $\lambda, ~\omega,~\theta$ are dimensionless. To create dimensionless parameters one defines $\tilde{G} =Gk^2$ and $\tilde{\Lambda} = \Lambda_{red} k^{-2}$.

The evolution of the curvature-squared coefficients is exactly the same as was previously calculated in dimensional regularization \cite{Julve:1978xn, Fradkin:1981iu}.
\begin{eqnarray}
\begin{aligned} \beta_{\lambda} &=-\frac{1}{(4 \pi)^{2}} \frac{133}{10} \lambda^{2} \\ \beta_{\omega} &=-\frac{1}{(4 \pi)^{2}} \frac{25+1098 \omega+200 \omega^{2}}{60} \lambda \\ \beta_{\theta} &=\frac{1}{(4 \pi)^{2}} \frac{7(56-171 \theta)}{90} \lambda \end{aligned}
\end{eqnarray}
These run only logarithmically in the usual way. In particular, the coefficient of the Weyl-squared term is asymptotically free and runs logarithmically to zero. The coefficient $\omega$ runs to a fixed point $\omega_* =-0.023$. Note however that in this evaluation the coefficient of the $R^2$ term $\omega/3\lambda$ is also indicative of asymptotic freedom because $\lambda$ is asymptotically free.

The remaining two couplings have an evolution
\begin{eqnarray}\label{dimensionful}
\begin{array}{l}{\beta_{\tilde{\Lambda}}=-2 \tilde{\Lambda}+\frac{1}{(4 \pi)^{2}}\left[\frac{1+20 \omega^{2}}{256 \pi \tilde{G} \omega^{2}} \lambda^{2}+\frac{1+86 \omega+40 \omega^{2}}{12 \omega} \lambda \tilde{\Lambda}\right]-\frac{1+10 \omega^{2}}{64 \pi^{2} \omega} \lambda+\frac{2 \tilde{G}}{\pi}-q(\omega) \tilde{G} \tilde{\Lambda}} \\ {\beta_{\tilde{G}}=2 \tilde{G}-\frac{1}{(4 \pi)^{2}} \frac{3+26 \omega-40 \omega^{2}}{12 \omega} \lambda \tilde{G}-q(\omega) \tilde{G}^{2}}\end{array}
\end{eqnarray}
with $q(\omega) =(83+70\omega +8\omega^2)/18\pi$. The initial factor in each beta function ($\pm 2$) is due to the explicit factor of $k$ used to make the couplings dimensionless. The remaining are due to perturbative interactions and these need to be large in order to cancel the $\pm 2$ if the beta function is to vanish. These perturbative terms are not found in dimensional regularization because they require powers of the cutoff.

If we follow Ref. \cite{Codello:2006in} and set $\omega$ and $\lambda$ to their fixed point values, the flow can be solved exactly. Expressing the result in terms of the Newton constant $G$ and vacuum energy density $\Lambda_{vac 0} $ defined at $k=0$, one finds,
\begin{equation}
  G(k) = \frac{G}{1+\frac{Gk^2}{g_*}}
\end{equation}
with $g_*\approx 1.4$ and
\beq\label{Lambdavac}
\Lambda_{vac}(k) = \Lambda_{vac 0} - \frac{1}{16\pi^2} k^4
\eeq
The quartic $k$ dependence of $\Lambda_{vac}$ is particularly striking. Evaluated at LHC energies, it would imply
\beq
\Lambda_{vac}(10 {\rm TeV})\sim - 10^{14}\rho_N \sim - 10^{61} \Lambda_{vac 0}
\eeq
where $\rho_N$ is the density of the nucleus and $\Lambda_{vac 0} \sim(10^{-3} {\rm eV})^4$ is the present experimental vacuum energy. It is also notable that the vacuum energy itself does not run to a UV fixed point. It increases without bound, and only the rescaled value ${\tilde \Lambda }\sim\Lambda_{vac}(k)/k^4$ stays finite.

However, this dependence is $k^4$ is actually illusory when it comes to applications of this parameter. Recall that $\Lambda_{vac}(k=0)=\Lambda_{vac  0}$ is meant to describe the vacuum energy density with all quantum corrections included, and $\Lambda_{vac}(k)$ is meant to describe that parameter with only quantum effects above the scale $k$ included. This implies that when we use $\Lambda_{vac}(k)$ we need also to add in the quantum corrections below $k$. For the vacuum energy this is seen to be related to
\beq\label{vacenergy}
\int^{k} \frac{d^3p}{(2\pi)^3} ~ \frac12 \omega_p = \frac{4\pi}{(2\pi)^3}\int_0^k p^2 dp ~\frac12 p= \frac{1}{16\pi^2} k^4
\eeq
If we add this back into Eq. \ref{Lambdavac} we get the full vacuum energy\footnote{The apparently missing factor of 2 in Eq. \ref{vacenergy} - for the 2 graviton helicity states - appears to come from the fact Eq. \ref{vacenergy} involves a non-covariant cutoff, while the Wetterich equation is a (Euclidean) covariant treatment. See also Ref. \cite{Ossola:2003ku, Akhmedov:2002ts}. Nevertheless, the principle remains the same. I thank Roberto Percacci for this observation.}. The running value is seen to be the full value with the effects of the momentum scales up to $k$ removed.

Similar considerations apply for the running $G(k)$. When using $G(k)$ one is instructed to also add in the quantum corrections from scales $0$ up to $k$. When this is done, one obtains the full $G$, which is the measured value.

{The functions $G(k)$ and $\Lambda(k)$ by default ``run'' because they depend on the scale $k$. However, we will see in the next section that they do not behave as gravitational running couplings in the sense of Weinberg, because they do not apply to physical processes. We will also explain the reason for this. Instead,} $\Lambda(k) $ and $G(k)$ are {\em incomplete} coupling constants. From their definition they include physics above the cutoff scale but not below. Indeed, insights from effective field theory indicate that the lower energy physics is the region that is dynamically important. Because of the uncertainty principle, physics from high energy scales beyond the active scale $k$ appears as local effects, parameterized by coefficients in a local action. Low energy physics can influence those local coefficients also (such that the cutoff scale disappears from physical observables) but also include dynamical effects from low energy propagation. The momentum dependence that we will see in the reactions to be described in Sect. \ref{explicit} all comes from low energy, as the high energy effects are only seen in the occasional unknown coefficient, such as $d_1$ in Eq. \ref{higherorder}. Because they are incomplete, parameters such as  $\Lambda(k) $ and $G(k)$ do not know about this low energy physics, and it is therefore not surprising that they do not capture the quantum physics seen in physical observables.

The AS running is an iterated one-loop calculation. The renormalization group is used to iterate the the matching at the scale $k$, which is itself performed at one loop order. For example, the full program has been performed in the quadratic truncation approximation of this section in Ref. \cite{Benedetti:2009rx}. This is an appropriate way to improve on the one-loop result of Codello and Percacci, but it does not change the fundamental interpretation of the cutoff dependence.

\section{The case against a running $G_N$ and $\Lambda$}

Quantum corrections and matter effects will clearly modify the physical value of $G$ and of the other parameters. However it is not a requirement that these organize themselves in a functional form that is usefully described by a running coupling. We can look at observables to see if this is the case.

The function $G(k)$ is defined to include all of the quantum effects above the cutoff scale $k$. In principle, it is designed to be supplemented by including all of the quantum effects {\em below} the scale $k$ also when using it to calculate some observable. The matching scale $k$ is unphysical and should drop out from physical observables once all quantum effects are included. Nevertheless, it is common AS practice to use $G_k$ as if it were the effective Newton constant at an energy of order $k$. However, one can see by direct calculations that this is not the case \cite{Anber:2011ut}. {The attempt to compare the form of $G(k)$ to low energy results is a valid test because the FRG predicts not only a UV fixed point but also the approach to the fixed point at lower energies with effective field theory calculations are performed. The same techniques which predict the fixed point also predict running at lower energies which overlaps with the validity of the EFT calculations. }

\subsection{Explicit calculations}\label{explicit}

Let us start by listing a series of physical amplitudes which have been calculated to one loop order. All of these have been calculated with the assumption that the value of the cosmological constant at low energy can be neglected. The results are then functions of $G$ and in some, but not all, cases contain coupling constants which are equivalent to a four-derivative truncation of the effective action. These reactions are observables. The question is whether we can define a useful running $G$ from these observables.

The most elemental quantum gravity process is the scattering of two gravitons. The lowest order scattering amplitude involves a
large number of individual tree diagrams but is given by the simple form
\begin{equation}
{A}^{tree}(++;++) = i\frac{\kappa^2}{4} \frac{s^3}{tu}\,,
\label{gravtree}
\end{equation}
where the signs $+,-$ refer to helicity indices and $s,t,u$ are the usual Mandelstam variables. In power counting, this is a dimensionless
amplitude of order $GE^2$. This was calculated at one loop order with the result.
The one loop amplitudes have been calculated by Dunbar and Norridge \cite{Dunbar:1994bn}. These are of order $G^2E^4$ and take the form

\begin{eqnarray}\label{eq:2}
{\cal A}^{1-loop}(++;--) & = & -i\,\frac{\kappa^4}{30720
\pi^2}
\left( s^2+t^2 + u^2 \right) \,,  \nonumber\\
{\cal A}^{1-loop}(++;+-) & = & -{1 \over 3}
{\cal A}^{1-loop}(++;--)\nonumber \\
{\cal A}^{1-loop}(++;++) & = &\frac{\kappa^2}
{4(4\pi)^{2-\epsilon}}\,
 \frac{\Gamma^2(1-\epsilon)\Gamma(1+\epsilon)}
 {\Gamma(1-2\epsilon)}\,
 {\cal A}^{tree}(++;++)\,\times(s\,t\,u)\\
&&\hspace{-0em}\times\left[\rule{0pt}{4.5ex}\right.
\frac{2}{\epsilon}\left(
\frac{\ln(-u)}{st}\,+\,\frac{\ln(-t)}{su}\,+\,\frac{\ln(-s)}{tu}
\right)+\,\frac{1}{s^2}\,f\left(\frac{-t}{s},\frac{-u}{s}\right)
\nonumber\\&&\hspace{1.4em}
+2\,\left(\frac{\ln(-u)\ln(-s)}{su}\,+\,\frac{\ln(-t)\ln(-s)}{tu}\,+\,
\frac{\ln(-t)\ln(-s)}{ts}\right)
\left.\rule{0pt}{4.5ex}\right]\,,\nonumber
\end{eqnarray}
where
\begin{eqnarray}\label{eq:f}
f\left(\frac{-t}{s},\frac{-u}{s}\right)&=&
\frac{(t+2u)(2t+u)\left(2t^4+2t^3u-t^2u^2+2tu^3+2u^4\right)}
{s^6}
\left(\ln^2\frac{t}{u}+\pi^2\right)\nonumber\\&&
+\frac{(t-u)\left(341t^4+1609t^3u+2566t^2u^2+1609tu^3+
341u^4\right)}
{30s^5}\ln\frac{t}{u}\nonumber\\&&
+\frac{1922t^4+9143t^3u+14622t^2u^2+9143tu^3+1922u^4}
{180s^4}\,.
\end{eqnarray}
Other amplitudes can be obtained from these by crossing.
I have discarded some purely infrared effects, including the expected IR radiative divergence. As noted by `t Hooft and Veltman, this reaction and all pure graviton processes will be independent of any coupling constants other than $G$ at this order, because the possible terms in the action vanish by the equations of motion $R_{\mu\nu}=0$.

Another core process is the gravitational potential for heavy masses. Including the leading quantum correction the potential has the form
\begin{equation}
V(r) = -G\frac{Mm}{r}\left[1+ \frac{41}{10\pi}\frac{G}{r^2}\right]\,,
\end{equation}
This particular definition is derived from the low energy limit of the scattering amplitude. I have dropped the leading classical correction. The quantum correction is universal, independent of the spin of the heavy particles.

The bending of light around a massive object can also be reliable calculated \cite{Bjerrum-Bohr:2014zsa, Bai:2016ivl, Chi:2019owc}.
\begin{equation}\label{e:theta}
\theta  \simeq {4G_N M\over b}+{15\over 4} {G_N^2 M^2 \pi \over
  b^2}
+\left(8 bu^S-47-64 \log {b\over 2b_0}\right)\,{\hbar G_N^2 M\over  \pi b^3}+\ldots\,.
\end{equation}
Here $1/b_0$ in the logarithm is the infrared cutoff
which removes the IR singularities of the amplitude.
Here there is not a universal behavior. The coefficient $bu^S$ is a parameter which depends on the intrinsic spin of the particle. It has values $371/120,~113/120, -29/8$ for scalars, the photon and the graviton respectively.

Dunbar and Norridge have also calculated the gravitational scattering of a massless scalar particle, $\phi +\phi \to \phi +\phi$ \cite{Dunbar:1995ed}.
At tree level, this has the form.
\begin{eqnarray}
{\cal M}_{tree}=i\frac{\kappa^2}{4}\left[\frac{st}{u}+\frac{su}{t}+\frac{tu}{s} \right]\,.
\end{eqnarray}
with as usual $\kappa^2 =32\pi G$. In this process there is a higher order operator which is needed to absorb the divergences which arise at one loop. This is
\begin{equation}\label{higherorder}
{\cal L}_2 = d_1 (D_\mu \phi D^\mu \phi)^2
\end{equation}
Including the renormalization of this higher order operator, the one loop hard amplitude is
\begin{eqnarray}
\nonumber
{\cal M}_{h}&=&i\frac{\kappa^4}{\left(4\pi\right)^2}\left\{\frac{(s^4+t^4)}{8st}\ln(-s)\ln(-t)+\frac{(s^4+u^4)}{8su}\ln(-s)\ln(-u)+\frac{(u^4+t^4)}{8tu}\ln(-t)\ln(-u) \right.\\
\nonumber
&&\left.\quad\quad\quad\quad+\frac{(s^2+2t^2+2u^2)}{16}\ln^2(-s) +\frac{(t^2+2s^2+2u^2)}{16}\ln^2(-t)+\frac{(u^2+2t^2+2s^2)}{16}\ln^2(-u)  \right.\\
\nonumber
&&\left.\quad\quad\quad\quad+\frac{1}{16}\left(\frac{st}{u}+\frac{tu}{s}+\frac{us}{t}  \right)\left(s\ln^2(-s)+t\ln^2(-t)+u\ln^2(-u)  \right)\right.\\
\nonumber
&&\left.\quad\quad\quad\quad+\left[ -\frac{(163u^2+163t^2+43tu)}{960}\ln\left(\frac{-s}{\mu}\right)-\frac{(163u^2+163s^2+43us)}{960}\ln\left(\frac{-t}{\mu}\right)\right.\right.\\
\label{AAAAhardScattering}
&&\left.\left.\quad\quad\quad\quad-\frac{(163s^2+163t^2+43ts)}{960}\ln\left(\frac{-u}{\mu}\right)+d_1^{ren}(\mu)(s^2+t^2+u^2) \right]\right\}\,,
\end{eqnarray}
where $\mu$ is an infrared scale. Again a purely infrared effect has been removed.

Anber and I have used the Dunbar-Norridge method to find the amplitudes for two different species of particles \cite{Anber:2011ut}. In the reaction $A+B \to A+B$  we find that the hard amplitude is
\begin{eqnarray}
\nonumber
{\cal M}_{h}&=&i\frac{\kappa^4}{\left(4\pi \right)^2}\left[\frac{1}{8}\left(s^4\ln(-s)\ln(-t)+u^4 \ln(-u)\ln(-t)\right) -\frac{1}{16t}\left(s^3+u^3+tsu \right)\ln(-t)+\frac{1}{16}\left(s^2\ln^2(-s)+u^2\ln^2(-u)\right)     \right.\\
\nonumber
&&\left.\quad\quad\quad+\frac{us}{16t}\left(s\ln^2(-s)+t\ln^2(-t)+u\ln^2(-u)  \right)+\frac{1}{240}\left(71us-11t^2\right)\ln(-t)-\frac{1}{16}\left(s^2\ln(-s)+u^2\ln(-u)\right) \right]\,,\\
\end{eqnarray}
For the crossed process, $A+{\bar A} \to B+{\bar B} $, one exchanges $s\leftrightarrow t$, which yields a significantly different functional form.

It is easy to see by inspection that there are no common factors for the power-law corrections to these processes. This is an immediate indication that there will not be a useful definition of a running $G$ which is useful in all processes. This is not a surprise as these kinematic effects do not amount to a direct renormalization of $G$. However, we can still proceed with an attempt to define a renormalization of $G$ at a higher renormalization scale $\mu_R$ and look at the outcome.

First consider graviton-graviton scattering. If we wish to renormalize this at high energy, we would like a kinematic configuration where all the kinematic variables are of the same large energy. In this case, we chose the central physical point $s =2E^2,~ t=u=-E^2$. If we use the amplitude ${\cal A}(++;++)$ and use this point to determine $G(E)$, we find
\begin{equation}
G^2(E) = G^2\left[1 +\frac{\kappa^2 E^2\left(\ln^2 2+\frac{1}{8}\left(\frac{2297}{180}+\frac{63\pi^2}{64}   \right)  \right)}{8\pi^2}\right]~~~.
\label{running1}
\end{equation}
We see that this definition leads to a growing running coupling $G(E)$, as opposed to the expectation from asymptotic safety of a
decrease in strength at high energy. Of course, since we are here using perturbation theory, we only should be obtaining the first order term in the expansion. Nevertheless the disagreement on the sign is clear.

We could alternatively consider the crossed reaction ${\cal A}(+,-;+,-)$ which is obtained from ${\cal A}(+,+;+,+)$ by the exchange $s\leftrightarrow t$. This makes the quantum corrections somewhat different, with the corresponding kinematic factor being
\begin{equation}
1 + \frac{\kappa^2 t}{16 \pi^2}
\left[ \ln\frac{-s}{t}\ln\frac{-u}{t}+ \frac{su}{2t^2} f \left(\frac{-s}{t},\frac{-u}{t}\right) \right]
= 1 +\frac{\kappa^2E^2\left(\frac{29}{10}\ln 2-\frac{67}{45}  \right)}{16\pi^2}
\end{equation}
instead of the factor in Eq. \ref{running1}.

If we used identical scalar particle scattering at the same kinematic point to identify a running coupling the result would be
\begin{equation}
G(E) = G\left[1-\frac{\kappa^2E^2}{360\left(4\pi\right)^2} \left(609\ln\frac{E^2}{\mu^2}+\left(340\pi^2+\left(123-340\ln 2\right)\ln2  \right)  \right)   \right]\,.
\label{running2}
\end{equation}
The single log term which appears in Eq. \ref{running2} could reasonably be associated with the higher order operator $d_1$, and perhaps should be removed from this expression.
Using the scattering of non-identical particles, one would find for   $A+B\to A+B$,
\begin{eqnarray}
{\cal M}_{total}=\frac{i\kappa^2E^2}{2}\left[1-\frac{\kappa^2 E^2}{10(4\pi)^2}\left(\left(19+10\ln 2\right)\ln\left(\frac{E^2}{\mu^2}\right)+5\left(\pi^2-(\ln 2-1)\ln 2  \right) \right)  \right]\,.
\end{eqnarray}
which would lead to yet a different running $G(E)$. On the other hand, using$A+{\bar A} \to B+{\bar B} $ we would have
\begin{eqnarray}
{\cal M}_{total}=\frac{i\kappa^2E^2}{8}\left[1+\frac{\kappa^2 E^2}{10(4\pi)^2}\left(9\ln\left(\frac{E^2}{\mu^2}\right)-5\pi^2+\left(19+5\ln 2  \right)\ln2 \right)  \right]\,.
\end{eqnarray}
The crossing problem is obvious here.

There is not much point to continue. It is clear that any application to other processes will yield yet other discordant results.  Even if we have an operational definition of a running $G$ at a higher renormalization point in one process, this definition does not apply to other reactions. This is not surprising, as the  quantum corrections here are not related to a renormalization of $G$.

We note also that having set the cosmological constant to zero at low energy, it stays zero in the scattering amplitudes. All the corrections come in at higher powers in the energy, in accord with the power counting theorems of the effective field theory.  The cosmological constant also does not run in these scattering amplitudes.

The examples here are evidence{ that the Weinberg criterion for AS is false}, {\em as applied to the parameters $\Lambda$ and $G$.} Even if we do not attempt to use the FRG form of the running $G$, there is no other form that does the job either. Nature does not organize itself like that at low energy. {Perhaps a revision of the Weinberg criterion is possible in which other parameters more important to the high energy limit have the flow envisioned by Weinberg.}

It is possible that in one given process - say, FLRW cosmology for example - it could be useful to define power-law running parameters for use in that setting and those running parameters might asymptote to an non-trivial UV fixed point. However, even if this is the case it would not imply that this defines a consistent quantum field theory of gravity. Such a field theory would have to be broadly applicable to all observables, and we have seen above a broad class of observables which do not share a useful running $G$.

\subsection{The driving force of the tadpole graph}

\begin{figure}[htb]
\begin{center}
\includegraphics[height=36mm,width=140mm]{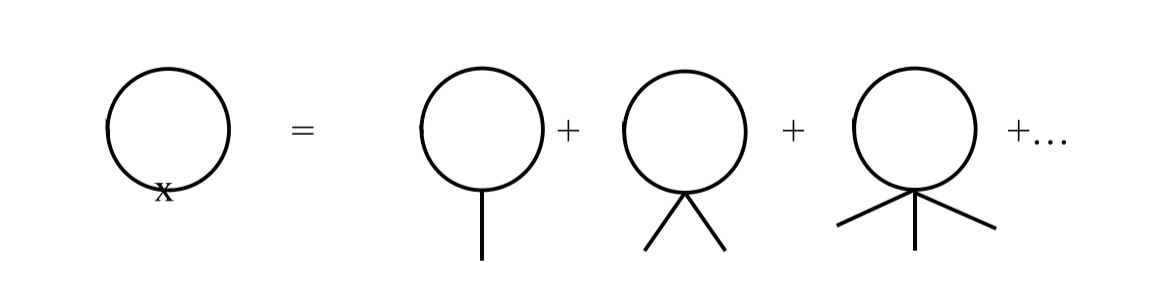}
\caption{The tadpole diagram on the left has an insertion of an operator involving the background field. When applied, this operator is expanded in powers of the external field, as on the right-hand side. The momenta of the external fields do not flow through the loop.}
\label{tadpole}
\end{center}
\end{figure}

We can look beyond the formalism and identify what is going wrong in the functional RG approach to the running $G$. The diagram driving the flow for this operator is the tadpole diagram of Fig. \ref{tadpole}. This diagram vanishes in dimensional regularization for massless particles. It is non-vanishing when evaluated with a cutoff. The issue is not really whether it vanishes or not, but that is a symptom. Since physical processes {\em can} be regularized dimensionally, we should not be surprised that there is not a signal of this diagram in the physical amplitudes. The more important feature is that this diagram does not feel the values of the external momenta, and here cutoff and dimensional regularization agree. Even with a cutoff, there is no external momentum flowing in loop. This tells us that the diagram does not know about the momentum scales of the physical reactions, and so cannot correspond to the use of running coupling depending on those scales. Once we identify how to treat this diagram, we will be able to bring the cutoff regularized result into agreement with dimensional regularization. To demonstrate this we need to look at the physics of the background field method.

With background field methods, one can capture the quantum effects using the heat kernel \cite{DeWitt:1967ub, Birrell:1982ix, Gilkey:1975iq, Donoghue:1992dd, Barvinsky:1985an, Codello:2012kq}, defined as
\beq
H(x,\tau) = <x|e^{-\tau {\cal D}}|x>
\eeq
for some differential operator ${\cal D}$. For example the functional determinant can be evaluated using
\beq
\Delta S = \int d^4x Tr <x|\log {\cal D}|x>
\eeq
with
\beq
<x|\log {\cal D}|x>=-\int_0^\infty \frac{d \tau}{\tau} <x|e^{-\tau {\cal D}}|x>  ~+~C
\eeq
The local heat kernel is expanded in powers of $\tau$ with the Seeley-DeWitt coefficients $a_i$, with the result
\beq
H(x,\tau) = \frac{i}{(4\pi )^{d/2}} \frac{e^{-\tau m^2}}{\tau^{d/2}}\left[a_0 (x) +a_1(x)\tau +a_2(x)\tau^2 +...\right]
\eeq
in an arbitrary dimension $d$. The contribution to the action is then
\beq
<x|\log {\cal D}|x> =\frac{-i}{(4\pi)^{d/2}}\left[ m^d \Gamma(-d/2) a_0(x) + m^{d-2}\Gamma(1-d/2) a_1(x)+ m^{d-4}\Gamma(2-d/2) a_2 (x) +...\right]
\eeq
As an example which is simpler than the graviton itself consider a scalar coupled to gravity with the Lagrangian
\beq\label{diffop}
\sqrt{-g}{\cal L} = \sqrt{-g}\frac12 \left[g^{\mu\nu}\partial_\mu \phi \partial_\nu\phi -m^2\phi^2 \right]
\eeq
in which the coefficients have the form
\begin{eqnarray}
a_0(x) &=& 1  \nonumber \\
a_1(x) &=& \frac16 R \nonumber \\
a_2(x) &=& \frac1{180}R_{\mu\nu\alpha\beta}R^{\mu\nu\alpha\beta}-\frac1{180}R_{\mu\nu}R^{\mu\nu} +\frac1{72}R^2
\end{eqnarray}
From this we see that $a_0$ is associated with the cosmological constant, $a_1$ is associated with the renormalization of $G$ and $a_2$ is asssociated with curvature-squared terms. In the AS beta functions this dependence is convoluted with the influence of the cutoff function, but this association remains true. I have included both a mass and a dimension $d$ in order to make the following points. In dimensional regularization for the massless graviton, we would set $m=0$ and the coefficients of $a_0$ and $a_1$ would vanish. The divergence in the coefficient $a_2$ is non-vanishing in the massless limit and is the usual divergence that one finds at one loop order. But also, in this evaluation the mass $m$ serves as a proxy for the IR cutoff of AS, with $m^2\sim k^2$. So we see that the $k^4$ and $k^2$ dependence of the running couplings comes form the $a_0$ and $a_1$ coefficients respectively.

{In 4D flat space, the Passarino-Veltman theorem \cite{Passarino:1978jh} says that all one loop diagrams can be reduced to scalar tadpole, bubble, triangle and box diagrams. The ``scalar'' part of this statement says that any momentum factors in the numerator can be removed and replaced by external momenta, leaving behind only the tadpole, etc diagrams with no momenta in the numerator. The heat kernel performs this operation describing the result using derivatives in the local operators, in our case $R,~R^2,~R_{\mu\nu}R^{\mu\nu}$ etc. The scalar tadpole, bubble, etc diagrams then contribute to the coefficients of the local operators. Each is readily identifiable by its dimension and divergence structure. In particular, in 4D the scalar tadpole has dimension $E^2$ and the scalar bubble is dimensionless, which is why they carry the $k^2$ and $\log k$ cutoff dependence. In curved spacetime, the use of the equivalence principle means that the short distance behavior of loops is equivalent to that of flat space. The use of Reimann normal coordinates can be used to describe the heat kernel and the AS RG flow using the same classification of tadpole, bubble, etc diagrams \cite{Codello:2012kq} including the non-local components of the heat kernel. The $k^2$ cutoff dependence of the $a_1$ coefficient is characteristic of the scalar tadpole diagram.    }

We can learn a bit more by looking at the ingredients to these heat kernel coefficients. Working in flat space for simplicity, we consider the differential operator as
\begin{eqnarray}
\log {\cal D} &=& \log [d_\mu d^\mu +m^2 +\sigma(x)] = \log [ \Box + m^2 + V(x)] \nonumber  \\
&=& \log\left[(\Box+m^2)(1+ \frac{1}{\Box+m^2}V)\right] \nonumber \\
&=& \log  [\Box +m^2]  + \frac{1}{\Box+m^2}V +\frac12 \frac{1}{\Box+m^2}V\frac{1}{\Box+m^2}V +....
\end{eqnarray}
where $d_\mu =\partial_\mu + \Gamma_\mu(x)$ and $\sigma(x) $ describe some interactions. Inserting a set of momentum eigenstates, we see that the first two terms in the heat kernel expansion are tadpole loops
\beq
\sim \int \frac{d^4p}{(2\pi)^4}  \times  \log  [\Box +m^2] \sim \int d m^2\int \frac{d^4p}{(2\pi)^4} \frac{1}{p^2 +m^2}
\eeq
and
\beq
 \sim \int \frac{d^4p}{(2\pi)^4} \frac{1}{p^2 +m^2}\times  V(x)
\eeq
These two are represented in Fig. 1. The key point here is that the tadpole has no external momenta flowing in these loops. This implies that when matrix elements are taken of the resulting effective Lagrangian, there will be no external momentum dependence coming from the $a_0$ and $a_1$ coefficients. This is already evident in the discussion of the one-loop running contributions to $\Lambda$ and $G$ in Sect. \ref{oneloop}. In contrast, the $a_2$ term is given by a bubble diagram, with two vertices and two propagators. It does involve the external momenta because it involves the interaction $V$ at different spacetime points. In addition to the local divergence which is contained in $a_2$ there is a non-local $\log q^2$ dependence. This can also be identified by a non-local version of the heat kernel method \cite{Codello:2012kq, Barvinsky:1985an}.

Combined with the discussion of Sect. \ref{oneloop}, we arrive at an understanding of how the cut-off regularization can agree with dimensional regularization. The dimensional regularization case integrates over all momenta with no separation of scales. The result is that the physical values of $\Lambda_{vac}$ and $G$ are not modified. In the cutoff regularization case, the so-called running couplings of $\Lambda(k)$ and $G(k)$ represent these parameters with quantum effects only above the scale $k$ included. They are actually incomplete couplings, where the the physics below the scale $k$ is missing. Technically, they are described by the tadpole diagram in which no momentum flows.  When supplemented by the rest of the loop below $k$ we again get the physical values of the parameters as the dependence on the separation scale must vanish. There is no external momentum flowing through these loops so that there is no net effect on the kinematic features of scattering amplitudes. This confirms that the $k$ dependence in $G(k)$ does not correspond to running in any kinematic sense. In contrast, the bubble diagram, associated with $a_2$ will contain logarithmic momentum dependence. Both dimensional regularization and cutoff regularization will agree on this and logarithmically running couplings associated with the $a_2$ coefficient will be physical.

\section{Comparison with quadratic gravity}\label{quadratic}

In this section, I discuss the AS result for the truncation including terms of order curvature squared, summarized above in Section \ref{oneloop}, with work on quadratic gravity, which uses the same operator basis but which does not use the AS machinery.

There are three points to be made in this comparison. 1) At least at one loop, this AS truncation is unsatisfactory in that when continued to Lorentzian spaces it contains a tachyon. It also contains a ghost state and violates causality on short time scales, although these may be less disastrous. 2) Further analysis of the ghost state indicates that there is an obstruction to the continuation from Euclidean space to Minkowski space, as there is a pole in the upper right quadrant of the complex $q_0$ plane. These are both problems that could could be due to the specific truncation, but which could in principle surface at any order of truncation in AS. 3) The third point is more positive: A focus on higher order terms in the graviton propagator may be useful for a Lorentzian variant of Asymptotic Safety.

\subsection{Tachyons and ghosts}\label{tachyons}

Because there are higher order terms in the most general action, the gravitational propagator will contain higher powers of $q^2$. With a truncation at order of the curvature-squared, this implies terms up to $q^4$ in the propagator. Normally these are forbidden by the K\"{a}llen-Lehmann representation of the propagator,
\beq
 D(q) =\frac{1}{\pi}\int_{4m_f^2}^\infty ds \frac{\rho(s)}{q^2-s+i\epsilon}
\eeq
with the spectral function $\rho(s)$ being positive definite,
which says that the propagator can fall by at most $q^{-2}$ at high momentum\footnote{There is the caveat that the KL representation does not necessarily apply to gauge-variant fields because the spectral function then does not correspond to the insertion of physical states.} . It then becomes clear that some of the usual assumptions of QFT (which forms the basis of the KL representation) must be given up in Asymptotic Safety (also in quadratic gravity). Some of the dangers are evidenced in the partial fraction decomposition of the propagator
\beq\label{quarticpropagator}
 iD(q) = \frac{i}{q^2- aq^4/M^2}= \frac{i}{q^2} - \frac{i}{q^2-M^2/a} \ \ .
\eeq
Here, $M$ is the intrinsic scale of the higher order terms, and I have included a parameter $a=\pm 1$ because the higher order behavior can come with either sign. For both signs of $a$, the second term in the partial fraction decomposition automatically comes with the ``wrong'' overall sign - it is a ghost. For $a=-1$ the ghost is also tachyonic in that it occurs for spacelike values of the four-momenta\footnote{Reminder: my metric convention is $(+,-,-,-)$.}. As far as I know, there is no way to rescue this situation. It leads to an unstable state with runaway production of tachyons. The $a=+1$ ghost is non-traditional in QFT, but seems to be more manageable. When treated properly, it can lead to a unitary theory \cite{Donoghue:2019fcb}, but one which violate microcausality \cite{Grinstein:2008bg, Donoghue:2019ecz}. However, these options are ones which any truncation of AS will be forced to confront.

The parameters of the one-loop AS solution given in Sec. \ref{oneloop} imply a tachyon in the spin-zero propagator and a $a=+1$ ghost in the spin-two propagator. Let us defer the discussion of the spin-two ghost to the next subsection. The spin-zero tachyon is a serious problem if it were to survive at higher order truncations. There is a bit of history/physics to understand concerning the tachyon. The first ingredient is that in this case, the high mass state is not ghost-like. It is the massless pole in the spin-zero channel which is ghost-like. That is, instead of Eq. \ref{quarticpropagator}, one has an overall minus sign,
\beq\label{quarticpropagator2}
 iD_0(q) = \frac{-i}{q^2- aq^4/M^2}= \frac{-i}{q^2} + \frac{i}{q^2-M^2/a} \ \ .
\eeq
That the massless pole is ghost-like is acceptable because the massless spin-zero component can be shown to be a gauge artifact \cite{Alvarez-Gaume:16}. The historical aspect is that several early works on the renormalization of quadratic gravity use what is now recognized to be the ``wrong'' sign without recognizing that this lead to tachyons. Adopting a modern parameterization for the quadratic terms, we have
\begin{equation}\label{quadraticorder}
S = \int d^4x \sqrt{-g}
\left[\frac{1}{6 f_0^2} R^2 - \frac{1}{2f_2^2} C_{\mu\nu\alpha\beta}C^{\mu\nu\alpha\beta} \right]
\end{equation}
in Lorentzian space. These signs lead to a normal massive spin-zero state, and the $a=+1$ spin two ghosts. Early work used the opposite sign on the $1/6f_0^2$ term, and concluded that both $f_0$ and $f_2$ are asymptotically free \cite{Julve:1978xn, Fradkin:1981iu}. With the non-tachyonic sign, $f_0$ is no longer asymptotically free \cite{Salvio:2014soa}. The Euclidean action of Sec. \ref{oneloop} shares yields asymptotic freedom for the overall $R^2$ coupling, and then would share the tachyonic property when continued to Lorentzian space.

It is possible that the tachyonic state could be removed using a higher order truncation {or no truncation at all. There are a few special functions whose Taylor expansion would show these poles when truncated at a fixed order, but which is well-behaved without the truncation.} However, this is already an indication that
simply obtaining a UV fixed point in the Euclidean FRG is not sufficient to claim that one has a well-behaved Lorentzian theory. Each
truncation must be checked separately. It is even more difficult to understand the ideal case, with no truncation.

\subsection{Obstacles to analytic continuation}\label{obstacles}

 The spin-two ghost in the quadratic truncation presents a more generic problem. There can be unexpected obstacles to the analytic continuation from Euclidean to Lorentzian spaces. {There has been some work on analytic continuation of the FRG in scalar theories \cite{Floerchinger:2011sc}, which however does not address the issue raised in this section.}

The location of the poles in the propagator has been explored in the quadratic gravity literature. I am particularly biased towards my own recent work with G. Menezes \cite{Donoghue:2019fcb,  Donoghue:2019ecz}, which is representative of the present status. The heavy ghost state will necessarily be unstable due to the coupling with the light gravitons and other light degrees of freedom. Including that coupling leads to a self-energy term in the propagator
\beq
i D_2(q) = \frac{i}{q^2+ \Sigma(q) - q^4/M^2}
\eeq
where $\Sigma (q) $ is the self energy. In gravity, there is a cut starting at $q^2=0$ where the self energy develops an imaginary part $\textrm{Im} ~\Sigma(q) = \gamma(q)$. Unitarity requires $\gamma(q) \geq 0$. The ghost resonance then has the form near $q^2= M^2$
\begin{eqnarray}  \label{merlin}
iD_2(q) &=& \frac{i}{q^2 - \frac{q^4}{M^2} + i \gamma(q) } \nonumber \\
&=& \frac{i}{\frac{q^2}{M^2}[M^2 - q^2 + i \gamma(q) (M^2/q^2)]} \nonumber  \\ &\sim& \frac{-i}{q^2-M^2 -i \gamma_M} \ \ .
\end{eqnarray}
This puts the resonance pole above the real axis
\beq
q^2 = M^2+i\gamma_M
\eeq
rather than usual resonances which occur below the real axis. In Ref \cite{Donoghue:2019ecz} we have labeled ghost resonances with this pole location as Merlin modes as they propagate backwards in time. We note that this construction would also work for higher order ghosts in the spin two channel. The fact that unitarity requires that $\gamma(q) \ge 0$, implies that all further ghost states would also live above the real axis.

For the purposes of quadriatic gravity, this is an arguably acceptable result. The resulting theory is unitary and stable near Minkowski space \cite{Donoghue:2019fcb}, but violates microcausality on timescales of order the width \cite{Donoghue:2019ecz, Grinstein:2008bg}, which is proportional to the inverse Planck scale. A look at the underlying calculations shows that this would appear to continue to happen if the propagator was defined with yet higher order dependence even if there were other unstable ghosts induced, as long as there were no tachyonic states allowed.
An AS theory defined in Lorentzian space would presumably share these acceptable features.

The danger for the present program of Asymptotic Safety is somewhat different. The original AS theory is defined in Euclidean space. To reach the real world, this needs to be continued to Lorentzian space. In amplitudes, this is accomplished by a rotation of the momentum space contour from the real axis to the imaginary axis, and is legitimate because there are no poles crossed by the rotation. The usual QFT rotation from Minkowski to Euclidean space is a tool which proves to be useful because of the usual analyticity properties of amplitudes. In the presence of higher derivatives, these analyticity properties are upset. This implies that there is no longer any guarantee that the Eucldean theory and the Minkowski theory share the same properties. The spin-two ghost found above is such a problem as would be any further ghosts.

{There has been recent work which attempts to keep the momentum dependence separate from the $k$ dependence and which addresses specific gravity amplitudes such as the propagator \cite{Christiansen:2012rx, Denz:2016qks, Falls:2018ylp, Bosma:2019aiu, Knorr:2019atm}.}
It appears that the spin-two ghost state is not just an artifact of the quadratic truncation. In a recent study by Bosma et al. \cite{Bosma:2019aiu},  the spin-two sector was parameterized much more generally,
\beq\label{formfactor}
C_{\mu\nu\alpha\beta} W(\Box) C^{\mu\nu\alpha\beta}
\eeq
where $W(\Box)$ is an arbitrary function, referred to as a form-factor. This directly impacts the spin-two propagator which becomes
\beq
iD_2(q) = \frac{i}{q^2-q^4 W(q^2)}
\eeq
Within the approximations of the calculation \cite{Bosma:2019aiu}, the result is approximated by
\beq\label{formfactor}
W(q^2)= w_\infty +\frac{\rho}{\alpha -q^2}
\eeq
where $\rho\simeq 0.015~ \alpha \simeq 1.8$  in Planck units and $w_\infty$ is a constant which is not determined by the calculation. In writing this result, I have made the continuation to Minkowski space in the most naive fashion - just changing the sign on the momentum. {The result in Ref. \cite{Bosma:2019aiu} is an approximate fit to the Euclidean numerical results and its full analytic structure is not precisely defined. Moreover,} the comments above about analytic continuation would also be applicable to this form-factor, and it is not clear how open channels would influence this continuation. In any case, this will have ghost poles when
\beq
q^2 W(q^2) =1
\eeq
Assuming that there are no tachyonic states, this {is still a} ghost pole. The form-factor description \cite{Bosma:2019aiu,  Knorr:2019atm} is a welcome new direction, because the functions of $\Box$ have direct physical relevance, in contrast with the unphysical parameter $k$.

\subsection{The graviton propagator and Lorentzian Asymptotic Safety}\label{alternative}

The higher order momentum dependence in the graviton propagator actually presents an opportunity for version of AS which is defined from the start in Lorentzian space. Potentially this could circumvent some of the problems which we have been discussing. However, it would require a reinterpretation of the program.

We have learned that low energy quantum effects involving $\Lambda$ and $G$ do not organize themselves in the way implied by present AS practice, or indeed of that suggested by the general Weinberg criterion.

However, we can also see that this may be irrelevant to the high energy behavior of the theory. In quadratic gravity, the propagator is modified by $q^4 $ terms, such that the effects of $\Lambda$ and $G$ ( of order $q^0$ and $q^2$) are sub-dominant at high energy, and the result is a renormalizeable theory. So the fact that there is not a good definition of a running $\Lambda$ and $G$ is not important for the overall structure of the theory. The parameters of the quadratic curvature terms are the essential ones for the renormalizablilty and running of the theory. In an AS framework, one could truncate at yet higher orders. This produces higher powers of momenta in the graviton propagator which are determine its high energy behavior.

Let us look at the potential for divergences in diagrams with these higher powers of the momenta. Consider the graviton propagator with the high energy behavior  $1/q^n$. For consistency, we need to keep vertices with powers of momentum running up to $q^n$, as the same operator which gives momentum dependence to the propagator will also give new vertices. The most divergent diagrams are the ones with the highest powers of momentum in the vertices, so we will consider that all vertices carry this maximal momentum factor. Let $N_{V}$ be the number of vertices, $N_I$ be the number of internal propagators, and $N_L$ be the number of loops. Then the overall high- momentum dependence of the diagram will be
\beq
(d^4q)^{N_L} ~(q^n)^{N_V} ~\frac{1}{(q^n)^{N_I}}
\eeq
from loop momenta, vertices and propagators\footnote{The factors of $q$ will in general involve external momenta, $q-p_i$ and after integration the amplitude will be expressed in terms of these $p_i$. Using dimensional regularization is useful here as it does not introduce extra dimensionful parameters, and the dimension in any divergence will be realized in terms of the external momenta. }. However, the number of internal propagators can be eliminated in favor of the number of vertices and loops. The relation is
\beq
N_I = N_L+N_V-1  \ \ .
\eeq
This converts the high energy behavior into
\beq
q^{D_n}  = (q)^{4 N_L} ~(q^n)^{N_V} ~\frac{1}{(q^n)^{N_L+N_V -1}}  = q^{(n + N_L(4-n))}
\eeq
which summarizes the divergence structure.

For two derivative actions, $n=2$ and we recover the well known power counting behavior of general relativity and chiral perturbation theory \cite{Weinberg:1978kz}
\beq
q^{D_2} =q^{(2+2N_L)}
\eeq
with tree level being $q^2$, one loop having divergences at $q^4$, two loop at $q^6$, etc. For $n=4$, such as for quadratic gravity, we recover power-counting renormalizability , with
\beq
q^{D_4}= q^4
\eeq
independent of the number of loops. For larger values of $n$ we get super-renormalizable behavior, with the diagrams becoming less divergent with higher loops. For example, for $n=6$, the power-counting gives
\beq
q^{D_6}= q^{6-2N_L}
\eeq
As the loop order increases, the amplitudes are increasingly focused on the infrared and are no longer divergent.
Phrased differently, tree-level amplitudes are always of order $q^n$ by assumption. For any $n$ there will be potential divergences at one loop order involving effects at order $q^4$. But then for larger $n>4$ the diagrams become more convergent at higher loop order.

This allows a possible reinterpretation of the AS program. Perhaps only some of the couplings need to be have the running behavior implied by the Weinberg criterion. Sub-dominant couplings such as $\Lambda$ and $G$ are not important for the program. {The important operators are those which dominate in the high energy limit. While there are in general there are an infinite number of these, the power counting above indicates that the damping provided by the higher powers of the graviton propagator may make a truncation at higher order feasible}. This inverts the present practice. Instead of a focus on low dimensional operators, one is more interested in higher dimensional operators that influence the graviton propagator. I note a similarity with the ``form-factors program'' \cite{Knorr:2019atm} in which the operators in the form factor, such as Eq. \ref{formfactor}, are higher powers of momentum in the graviton propagator. It would be interesting to see if this program could be formulated in Lorentzian spacetime.

Of course, this suggestion is still somewhat vague and needs to be better developed. One still needs to avoid tachyons and deal with ghosts. But it does point to a form of Asymptotic Safety that can be described from the start in Lorentzian spaces, and which can be in agreement with explicit calculations at low energy. Moreover, it is clear that the high momentum behavior of the graviton propagator is of special significance as it determines the UV properties of loop diagrams.

\section{Overall assessment}

We have examined in particular the running Newton constant $G(k)$ within AS and argued that it is not valid for use in the real world. The reasons for that include:

1) It does not capture the energy dependence in explicit observables. There are kinematic and universality obstacles to any such use. Note that these examples are also counter-examples to the Weinberg conditions for Asymptotic Safety if applied to $G,~\Lambda$. {If the Weinberg vision for Asymptotic Safety is to continue, the conditions need to be modified to exclude the low energy parameters $G, ~\Lambda$.}

2) The definition of the $G(k)$ and $\Lambda(k)$ are such that they include quantum effects beyond the scale $k$. They should be supplemented with the quantum effects below $k$. When this is done, the intermediate scale $k$ should disappear.

3) We can also see that the values of $G(k)$ and $\Lambda(k)$ arise from the tadpole diagram, which a) vanishes in dimensional regularization and b) does not contain any external momentum flow through the loop. This loop will not influence the kinematic behavior of reactions.

\noindent {Points 2 and 3 indicate that these couplings are what I have referred to as incomplete couplings rather than running couplings in the sense of the Weinberg criterion. They become complete only in the $k\to 0$ limit. In this sense there is a disconnect between essentially all of present AS practice and the {Weinberg conditions of Eq. \ref{initial} -\ref{fixed}}. It needs to be recognized that the cutoff dependence of $G(k)$, $\Lambda(k)$ and likely many of the higher power couplings is not the same as the running couplings in physical reactions. These features are most problematic in attempts to apply Asymptotic Safety in phenomenological settings. Some of the previous phenomenological applications have been discussed in the surveys of the subject \cite{Niedermaier:2006wt, Reuter:2019byg} .} The use of these couplings is not appropriate for phenomenological applications and does not satisfy the goals of Asymptotic Safety.

{In the process of making these comparison, it can be recognized that at least a portion of {Weinberg's conditions for} Asymptotic Safety fails at the energies which we have considered - that which applies to the proposed running of $G$ and $\Lambda$. Not only does the FRG version of running fail to match explicit calculations, but even operationally there is no form that will work at scales below $M_P$. Nature does not organize itself this way. This need not be a fatal flaw, as these couplings describe operators which are sub-dominant in the high energy limit. Higher powers of curvatures and derivatives will dominate at high energy, and so it is possible that even if $G$ and $\Lambda$ do not run, the important couplings at high energy do. This is what happens in quadratic gravity, where the curvature squared terms make the theory renormalizeable and their coefficients do have logarithmic running. However, there still needs to be a reformulation of the Weinberg criterion which takes into account the multiple kinematic variables of different magnitudes and signs which complicate to running of non-logaritmic power-law couplings.}

This leaves the ``ideal FRG program'' as a possibility. Here one integrates in Euclidean space down from the UV fixed point all the way to $k=0$. {The couplings have ``run'' in the theory space of coupling constants not in the real space of energies and momenta, and have completed their evolution by taking the $k\to 0$ limit. At intermediate values of $k$ these couplings are not considered to be physical, but their $k=0$ limit defines an action with an infinite number of terms, which is then to be applied in Lorentzian space. The action is described by an infinite number of parameters such as $G$ and $\Lambda$, which are themselves just constants defined by their $k\to 0$ limit. These couplings are correlated - fixed by a smaller number defined at the fixed point. This appears to be the situation advocated in Section 6.18 of the Wetterich review \cite{Wetterich:2019qzx}. However, it is a very different situation than the Asymptotic Safety envisioned by the Weinberg conditions in Eq. \ref{initial}-\ref{fixed}, where the running couplings were functions of energy applied in physical reactions.} Here I have raised two cautions:

1) Any truncation of this ideal action will have ghosts, and possibly tachyons. These have to be understood and managed.

2) Any truncation without tachyons will likely have one or more obstacles to the analytic continuation from Euclidean to Lorentzian space. These are poles in the graviton propagator that occur in the quadrants needed for the Euclidean rotation.

\noindent {There can be a significant difference between a Euclidean theory and a Lorentzian one in the presence of operators with higher derivatives/curvatures.}

It is possible that both of these points can be overcome. However, even if this occurs, we do not have any indication on why the resulting theory would satisfy the Weinberg criterion {or}lead to finite results in physical observables. {The Weinberg criterion gave an intuitive rationale for the finiteness of the theory. But if this ideal FRG program does not {generate running parameters in physical reactions}, we need a new rationale. If the cutoff dependence in $G(k)$ etc is not the same as the running of couplings in physical reactions, what reason do we have to expect that we get finite high energy limits for such reactions? The existence of a Euclidean UV fixed point is not sufficient by itself for this result. Indeed, existing truncations do not satisfy this despite all having such fixed points. One needs to obtain finite results for an infinite number of processes at an infinite number of kinematic points. One does have an infinite number of couplings, but the mechanism for success is unknown.}

On the more positive side, I have argued that maybe a Lorentzian version of AS could occur through a focus on the higher order terms contributing to the graviton propagator. The basic point here is that $\Lambda$ and $G$ become unimportant at high energy in the graviton propagator when higher powers of of $q^n$ appear in the propagator. This is seen in quadratic gravity where the inclusion of $q^4$ terms in the propagator lead to a renormalizeble theory, and is encountered in Euclidean AS through the inclusion of form-factors \cite{Knorr:2019atm} . I have used power counting to argue that one could perhaps get a Lorentzian theory with these higher order terms.

\section*{Acknowledgements}  This work has been supported in part by the National Science Foundation under grant NSF-PHY18-20675. I am particularly grateful to Mohamed Anber, Alessandro Codello and Roberto Percacci for discussions over the years, and also F. Saueressig, A. Eichhorn, M. Reuter, D. Litim, S. Weinberg among many others.

\end{document}